\documentclass[12pt]{iopart}

\usepackage{amssymb}
\usepackage{graphicx}
\usepackage{amsbsy}
\usepackage{float}
\usepackage{color}
\usepackage{bm}

\usepackage{marvosym}

\newcommand{\bq}{\begin{eqnarray}}
\newcommand{\eq}{\end{eqnarray}}
\newcommand{\bqn}{\begin{eqnarray*}}
\newcommand{\eqn}{\end{eqnarray*}}

\newcommand{\sss}{\mathbf{s}}
\newcommand{\qq}{\mathbf{q}}
\newcommand{\QQ}{\mathbf{Q}}
\newcommand{\dd}{\mathbf{d}}
\newcommand{\ee}{\mathbf{e}}
\newcommand{\pp}{\mathbf{p}}
\newcommand{\xxi}{\bm{\xi}}

\begin{document}
%%%%%%%%%%%%%%%%%%%%%%%%%%%%%%%%%%%%%%%%%%%%%%%%%%%%%%%%%%%%%%%%%%%%%%%%%%%%%%
%%%%%%%%%%%%%%%%%%%%%%%%%%%%%%%%%%%%%%%%%%%%%%%%%%%%%%%%%%%%%%%%%%%%%%%%%%%%%%
%%%%%%%%%%%%%%%%%%%%%%%%%%%%%%%%%%%%%%%%%%%%%%%%%%%%%%%%%%%%%%%%%%%%%%%%%%%%%%
%\journalname{Journal of Statistical Physics}
\title{The density of a fluid on a curved surface}
%\titlerunning{The density of a fluid}  
%\dedication{$\heartsuit$ to my daughter $\mathit{Alice}$ \Taurus} 
\author{Riccardo Fantoni}
%\authorrunning{R. Fantoni}
\ead{rfantoni@ts.infn.it}
%\author{Jeandrew Brinks}
%\email{jeandrew@sun.ac.za}
\address{National Institute for Theoretical Physics (NITheP) and
Institute of Theoretical Physics, University of
Stellenbosch, Stellenbosch 7600, South Africa} 
%\author{Gabriel T\'ellez}
%\email{gtellez@uniandes.edu.co}
%\affiliation{Grupo de F\'{\i}sica T\'eorica de la
%Materia Condensada, Departamento de F\'{\i}sica, Universidad de Los
%Andes, A.A. 4976, Bogot\'a, Colombia}
%\author{Karl M\"oller}
%\affiliation{Institute of Theoretical Physics, University of
%Stellenbosch, Stellenbosch 7600, South Africa}

%\author{Peter J. Forrester}
%\email{p.forrester@ms.unimelb.edu.au}
%\affiliation{Department of Mathematics and Statistics, The University
%of Melbourne, Victoria 3010, Australia}
%\author{Filippo Giraldi}
%\email{filgi@libero.it, giraldi@uknz.ac.za}
%\affiliation{Quantum Research Group, School of Physics and National
%Institute for Theoretical Physics, University of KwaZulu-Natal,
%Westville Campus, Durban 4001, South Africa.}

\date{\today}

\begin{abstract}
We discuss the property of the number density of a fluid of
particles living in a curved surface without boundaries to be constant
in the thermodynamic limit. In particular we find a sufficient
condition for the density to be constant along the Killing 
vector field generating a given isometry of the surface and the
relevant necessary condition. We reinterpret the effect of a
curvature on the fluid in a physical way as responsible of an external
``force'' acting on the particles.
\end{abstract}

%\keywords{Statistical physics; fluids; Riemannian surfaces; number
%  density; scalar curvature; isometry; Coulomb potential} 

\maketitle
%%%%%%%%%%%%%%%%%%%%%%%%%%%%%%%%%%%%%%%%%%%%%%%%%%%%%%%%%%%%%%%%%%%%%%%%%%%%%%
\section{Introduction}
%%%%%%%%%%%%%%%%%%%%%%%%%%%%%%%%%%%%%%%%%%%%%%%%%%%%%%%%%%%%%%%%%%%%%%%%%%%%%%
\label{sec:introduction}

The physics of fluids of particles living in surfaces is a well known
chapter of surface physics. A special role is played by low
dimensional exactly analytically solvable fluids, as they inform
approximate solutions in higher dimensions and 
general sum rules. In the statistical mechanics of continuous fluids,
those where the particles are allowed to move in a continuous space,
one finds exact  solutions for various Coulomb 
fluids \cite{March84}. For example the one-component Coulomb plasma
(OCP) is exactly solvable in one-dimension \cite{Edwards62}.
In two dimensions B. Jancovici and A. Alastuey
\cite{Jancovici81b,Alastuey81} proved that the OCP is exactly solvable
analytically at a special value of the coupling constant, in their 1981
work. Since then, a growing interest in two-dimensional plasmas has
lead to study this system on various flat geometries  
\cite{Rosinberg84,Jancovici94,Jancovici96} and two-dimensional curved
surfaces like the cylinder \cite{Choquard81,Choquard83}, 
the sphere 
\cite{Caillol81,Jancovici92,Jancovici96b,Tellez99,Jancovici00}, the
pseudosphere \cite{Jancovici98,Fantoni03jsp,Jancovici04}, and the
Flamm's paraboloid \cite{Fantoni2008,Fantoni2012}. Among these
surfaces only the 
last one is of non-constant curvature. The statistical mechanics of
liquids and fluids in curved spaces is a field of growing interest
\cite{Tarjus2011}. 

Here we do not restrict ourselves to those exactly solvable cases but
want to find a general property of any given fluid living on a curved
surface without boundaries. A {\sl homogeneous} fluid living in a
plane (or in general an Euclidean space) is known \cite{Hill} to have
a {\sl constant density}. This same conclusion holds for a
(non-ideal) fluid living in a surface of constant curvature in its
thermodynamic limit \footnote{The notion of thermodynamic limit will
  become clear further on in the paper.}. In this
paper we will state what can be said about the constancy of the
density for a fluid living in a Riemannian surface without boundaries
and embeddable in the three dimensional Euclidean space,
in its thermodynamic limit. It is obvious that an ideal fluid (a gas)
has a constant density on any surface and whether or not we are in the
thermodynamic limit. But what can be said about the non-ideal fluid? 

The study of Ref. \cite{Fantoni2008} showed that the OCP in the
Flamm's paraboloid is indeed homogeneous. We expect this
occurrence to be due to the long range nature of the Coulomb potential
and argue that it cannot hold in general for other choices of the pair
potential or of the surface. 

In this work we will give a physical interpretation to the curvature
of the surface as an external ``force'' guiding the particles of the
corresponding ``flat'' fluid. We will show that the Coulomb potential
has to be a function of the geodesic distance between the charges and
we will restrict to a definition of a fluid as one made of particles
with a pair interaction potential which is a function of the geodesic
distance between the two particles. We will then find a necessary and
sufficient condition for the density multiplied by the square root of
the determinant of the metric tensor to be constant along a certain
direction. We will show how this condition holds true both for
non-quantum and quantum fluids. 
 
The paper is organized as follows: in Sec. \ref{sec:problem} we state
the problem we want to solve at the level of the non-quantum fluids;
in Sec. \ref{sec:curvature} we reformulate the problem in such way as
to make explicit the physical interpretation of the curvature of the
surface; Sec. \ref{sec:quantum} is
devoted to the quantum fluid formulation of the problem;
Sec. \ref{sec:conclusions} is for final remarks. 
 
%%%%%%%%%%%%%%%%%%%%%%%%%%%%%%%%%%%%%%%%%%%%%%%%%%%%%%%%%%%%%%%%%%%%%%%%%%%%%%
\section{Statement of the problem}
%%%%%%%%%%%%%%%%%%%%%%%%%%%%%%%%%%%%%%%%%%%%%%%%%%%%%%%%%%%%%%%%%%%%%%%%%%%%%%
\label{sec:problem}

Given a non-quantum fluid of point wise particles {\sl living in} a
surface ${\cal S}$ 
embeddable in the three dimensional Euclidean space (note that we will
not take under consideration those surfaces deriving from a Riemannian
metric but not embeddable and those not deriving from a metric) and
without boundaries one can define the canonical ensemble particle {\sl
  number density} as 
\cite{Hill} 
\bq \label{rho}
\rho(\qq_1)&=&\frac{N}{Z}\int_{\Omega} e^{-\beta V(\qq_1,\ldots,\qq_N)}
\prod_{i=2}^N\sqrt{g(\qq_i)}\wedge_{\alpha_i=1}^2\dd q^{\alpha_i}~,\\
Z&=&\int_\Omega e^{-\beta V(\qq_1,\ldots,\qq_N)}
\prod_{i=1}^N\sqrt{g(\qq_i)}\wedge_{\alpha_i=1}^2\dd q^{\alpha_i}~,
\eq
where $N$ is the number of particles confined in the region $\Omega$,
$\beta=1/k_BT$ with $k_B$ Boltzmann's constant and $T$ the absolute
temperature. The potential energy of the fluid is
$V=\sum_{1\le i<j\le N}v(d(\qq_i,\qq_j))$ where
$v$ is the pair-potential and $d(\qq,\qq^\prime)$ is
the geodesic distance between the two points $\qq$ and
$\qq^\prime$. The surface is defined by a metric tensor 
$g_{\alpha\beta}$ so that the square of the proper length of the
infinitesimal line element is given, using the 
usual Einstein's summation convention, by
$\dd\sss^2=g_{\alpha\beta}(\qq)\dd q^\alpha\otimes\dd q^\beta$ where
$\otimes$ is the usual tensor product. We 
denote with $g(\qq)=\det||g_{\alpha\beta}(\qq)||$ the Jacobian of the
transformation from a locally flat reference frame to the local
coordinates system on the surface. Here we use a coordinate basis
$\{\ee_\alpha=\partial_{q^\alpha}\}$ so that
$\qq=q^\alpha\ee_\alpha$ and the symbol $\dd$ stands for the
exterior derivative. As usual we use upstairs Greek indexes for
contravariant components and downstairs Greek indexes for covariant
components, and we use a downstairs roman index to denote the
(distinguishable) particle number. The symbol $\wedge$ 
indicates the usual wedge product. In the following we will call
$\mbox{vol}(\Omega)=\int_\Omega\sqrt{g(\qq)}\wedge_{\alpha=1}^2\dd 
q^\alpha$ the {\sl volume} of the region $\Omega$. 

The problem we want to discuss is the one of finding continuous
transformations that leave unchanged the density $\rho(\qq)$ in the
{\sl thermodynamic limit}. Here we think of the surface ${\cal S}$ as
an embeddable one without boundaries. And by thermodynamic limit we
mean that if ${\cal S}$ extends to infinity, $\mbox{vol}(\Omega)\to
\infty$ with $\overline{\rho}=N/\mbox{vol}(\Omega)$ kept constant or
if ${\cal S}$ is closed, $\Omega\to {\cal S}$ with
$\overline{\rho}=N/\mbox{vol}(\cal S)$. We want to answer the
question: 
``when is $\rho(\qq)$ constant on ${\cal S}$ in the thermodynamic
limit?''. 

The number density satisfies the following normalization
condition
\bq \label{vol}
\int_\Omega\rho(\qq)\sqrt{g(\qq)}\wedge_{\alpha=1}^2\dd q^\alpha=
N=\mbox{vol}(\Omega)\overline{\rho}~.
\eq
So when the density is constant in the surface we must have
$\rho=\bar{\rho}$. 

%%%%%%%%%%%%%%%%%%%%%%%%%%%%%%%%%%%%%%%%%%%%%%%%%%%%%%%%%%%%%%%%%%%%%%%%%%%%%%
\section{Reinterpretation of the curvature}
%%%%%%%%%%%%%%%%%%%%%%%%%%%%%%%%%%%%%%%%%%%%%%%%%%%%%%%%%%%%%%%%%%%%%%%%%%%%%%
\label{sec:curvature}

Choosing the coordinate basis so that $\xxi=\partial_{q^\alpha}$ is a
Killing vector field \cite{MTW} generating an {\sl isometry}, then 
$g_{,\alpha}=0$, where we use the usual comma convention to indicate a
partial directional derivative. We know that if $\pp$ is the
momentum of a free particle on ${\cal S}$ then $\pp\cdot\xxi$ is a
constant of motion $p_\alpha(\pp\cdot\xxi)^{;\alpha}=0$, where we use
the usual semicolon convention to indicate a covariant derivative. The
ideal gas has constant density on every surface regardless of the
curvature and of the thermodynamic limit. We thus have to worry about
the term $\exp(-\beta V)$.  
Now, if one moves the $N$ particles at $\qq_1,\ldots,\qq_N$ along the
vector field $\xxi$ the geodesic
distances among the system of particles will stay constant as well as
the potential energy $V$. We then have proven that given a Killing
vector field $\partial_{q^\alpha}$ then $\rho_{,\alpha}=0$. Strictly
speaking before taking the thermodynamic limit, the domain 
has boundaries, and close to these, one might not be able to move the
particles along the Killing vector field, invalidating the conclusion
near the boundary. 
When taking the thermodynamic limit, one needs to be able to quantify
if these boundary effects will be negligible or not, and how deep they can
affect the bulk of the system. This depends of the pair potential $v$ and on
the surface. In the flat space it is well known that the boundary effects
are negligible (for suitable short-ranged potentials and for the Coulomb
potential for globally neutral systems to have screening). But for a general
curved surface, a proper study of what happens in thermodynamic limit
with this boundary effect is needed and it will certainly impose additional
conditions on the pair potential $v$, and probably also on the surface, to
keep valid the conclusion that $\rho_{,\alpha}=0$. The conditions on the surface
might appear for example in cases similar to the pseudosphere, where it
has been shown that boundary effects can be of the same order of
magnitude as the bulk properties (see 
Refs.\cite{Jancovici98,Fantoni03jsp,Jancovici04}). So, additional
work in this direction is needed. 

This is clearly only a {\sl sufficient} condition but it is enough to say
that on the sphere (or the plane), the surface of constant curvature
\cite{Wolfgang06}, where $\xxi=\partial_{\varphi}$, with
$\varphi$ the azimuthal angle, the density will be constant in the
thermodynamic limit. One, in fact, has that the density is constant along
parallels. And this, given the symmetries of the sphere, means that the
density is indeed everywhere constant over the whole sphere, with
$\rho=\overline{\rho}$. 

On the other hand a {\sl necessary} condition can be expressed as
follows: Say that we find a coordinate system such that, for all $v$,
$(\sqrt{g}\rho)_{,\alpha}=0$ then in particular for $v=0$ we have
$\rho=$constant and $g_{,\alpha}=0$. For the Flamm's paraboloid 
\cite{Fantoni2008} we can say that there certainly exists a fluid (at least
one $v$) such that $(\sqrt{g}\rho)_{,r}\neq 0$ since $\partial_{r}$ is
not a Killing vector of the surface and $g_{,r}\neq 0$. And we know
\cite{Fantoni2008} that the OCP is an example.  

The problem then reduces to understand what can be said about surfaces
of non-constant curvature. Note that we can as well rewrite
Eq. (\ref{rho}) as follows
\bq \label{rhog}
\sqrt{g(\qq_1)}\rho(\qq_1)=N\frac{\int_\Omega 
e^{-\beta [V(\qq_1,\ldots,\qq_N)+\sum_{i=1}^N\phi(\qq_i;\beta)]}
\prod_{i=2}^N\wedge_{\alpha_i=1}^2\dd q^{\alpha_i}}
{\int_\Omega e^{-\beta [V(\qq_1,\ldots,\qq_N)+\sum_{i=1}^N
\phi(\qq_i;\beta)]}
\prod_{i=1}^N\wedge_{\alpha_i=1}^2\dd q^{\alpha_i}}~,
\eq
where $\phi(\qq;\beta)=-[\ln g(\qq)]/2\beta$ is an ``external
potential''. A form which suggests, on physical grounds, a local
dependence of the density on the curvature. The fluid is seen in this
formulation as living on a ``flat space'', the two dimensional space
determined by the local coordinates chart $(q^1,q^2)$ used in the
surface, subject to an external potential induced by the metric. This
suggestive reinterpretation of the problem can sometimes lead to a
wrong intuition. For example we know that the
OCP on the Flamm's paraboloid (see Sec. 4.2.4
of Ref. \cite{Fantoni2008}) has a density that is everywhere
constant even if this surface is only asymptotically flat but curved
near the ``horizon'', the scalar curvature being proportional to the
Euclidean distance $r$ from the origin to the power of minus
three. Whereas the constancy of the density along the azimuthal
direction $\varphi$ has to be expected from the sufficient condition
stated above, the constancy of the density along the radial $r$
direction is not at all intuitive, even more so at the light of the
discussion which follows. 

For a surface with a conformal metric
$g_{\alpha\beta}=\sqrt{g(\qq)}\delta_{\alpha\beta}$, \footnote{Note
  that the following are all surfaces of this kind: the sphere
  embedded in three dimensional Euclidean space $\sqrt g=4/(1+s^2)^2$,
  the pseudosphere embedded in three dimensional Minkowski space
  $\sqrt g=4/(1-s^2)^2$, the cylinder embedded in three dimensional
  Euclidean space $\sqrt g=1$, and the Flamm's paraboloid embedded in
  three dimensional Euclidean space $\sqrt g=(1+1/s)^4$. Here
  $s=\sqrt{(q^1)^2+(q^2)^2}$.}
the scalar curvature $R$ can be written as  
\bq
R(\qq)=e^{\beta\phi(\qq)}\beta\Delta_{\mbox{\scriptsize flat}}\phi(\qq)~,
\eq
where $\Delta_{\mbox{\scriptsize flat}}=\partial^2_{q^1}+\partial^2_{q^2}$ is the flat
Laplace's operator. The external ``force'' acting on the particles due
to the curvature is then $-R\exp(-\beta\phi)/\beta$. For the Flamm's
paraboloid \cite{Fantoni2008} the force acting on the charges turns
out to be $4/[\beta s (1+s)^2]$ where $s$ $=$
$\sqrt{(q^1)^2+(q^2)^2}$. As we already
mentioned above, in this case, the OCP shows a constant
density in the surface. In Section \ref{sec:coulomb-fluid} we show
that in general it has to be expected a non-constant density.

On the other hand the formulation of Eq. (\ref{rhog}) suggests that
certainly $\sqrt{g}\rho$ is a more fundamental quantity than just
$\rho$ itself to look upon.
%%%%%%%%%%%%%%%%%%%%%%%%%%%%%%%%%%%%%%%%%%%%%%%%%%%%%%%%%%%%%%%%%%%%%%%%%%%%%%
\subsection{The Coulomb pair potential} 
%%%%%%%%%%%%%%%%%%%%%%%%%%%%%%%%%%%%%%%%%%%%%%%%%%%%%%%%%%%%%%%%%%%%%%%%%%%%%%
\label{sec:coulomb}

Here we want to show that the Coulomb potential between two charged
particles living in a given surface ${\cal S}$ has to be a function of
the geodesic distance between the charges
\cite{Jancovici81b}
\cite{Choquard81}
\cite{Caillol81}
\cite{Jancovici98,Fantoni03jsp} 
\cite{Fantoni2008}.

The Coulomb potential is defined by the Poisson's equation,
\bq \label{lgf}
\Delta_\qq v_{Coul}(\qq,\qq^\prime)=-2\pi
\delta^{(2)}(\qq,\qq^\prime)~,
\eq
where $\Delta_\qq$ is the Laplace-Beltrami operator and
$\delta^{(2)}(\qq,\qq^\prime)=\delta^{(2)}(d(\qq,\qq^\prime))$ the
Dirac delta function, in the surface ${\cal S}$. The Laplace-Beltrami
operator is invariant to isometries. This means that if the charge at
$\qq$ and the one at $\qq^\prime$ are moved along the vector field of
an isometry the Laplace-Beltrami operator will not change. Neglecting
eventual additive functions which have a null Laplacian we must have  
\bq \label{vcoulomb}
v_{Coul}=f(d(\qq,\qq^\prime))~.
\eq
For example on the sphere \cite{Caillol81} of radius $R$ one finds
$f(x)=-\ln (2R\sin(x/2R)/L)$ 
with $L$ a length scale. The conclusion of Eq. (\ref{vcoulomb}) is in
agreement with Fermat's principle for light propagation \cite{Rossi}.

%%%%%%%%%%%%%%%%%%%%%%%%%%%%%%%%%%%%%%%%%%%%%%%%%%%%%%%%%%%%%%%%%%%%%%%%%%%%%%
\subsection{The Coulomb fluid} 
%%%%%%%%%%%%%%%%%%%%%%%%%%%%%%%%%%%%%%%%%%%%%%%%%%%%%%%%%%%%%%%%%%%%%%%%%%%%%%
\label{sec:coulomb-fluid}

For an open surface with a conformal metric
$g_{\alpha\beta}=(\sqrt{g(s)}/s)\delta_{\alpha\beta}$, $s\in
[0,+\infty[$ the Laplace-Beltrami operator
can be rewritten as   
\bq
\Delta f=\frac{s}{\sqrt{g}}\Delta_{\mbox{\scriptsize flat}} f~,
\eq
where $\Delta_{\mbox{\scriptsize flat}}$ is the usual Laplace operator in flat space
$(x=s\cos\varphi, y=s\sin\varphi)$. We can then introduce a complex
coordinate $z=se^{i\varphi}$ and the Laplacian Green's function
(\ref{lgf})
\bq \label{lf}
\Delta_{\mbox{\scriptsize flat}}v_{Coul}((s,\varphi),(s_0\varphi_0))=
-2\pi\frac{1}{s}\delta(s-s_0)\delta(\varphi-\varphi_0)
\eq
 can be solved as usual, by using the decomposition as a
Fourier series. Since (\ref{lgf}) reduces to the flat Laplacian Green's
function, the solution is the standard one 
\bq \label{fs}
v_{Coul}((s,\varphi),(s_0\varphi_0))=\sum_{n=1}^\infty\frac{1}{n}
\left(\frac{s_<}{s_>}\right)^n\cos[n(\varphi-\varphi_0)]+v_0(s,s_0)~,
\eq
where $s_>=\mbox{max}(s,s_0)$ and $s_<=\mbox{min}(s,s_0)$. The Fourier
coefficient for $n=0$ has the form
\bq
v_0(s,s_0)=\left\{\begin{array}{ll}
a_0^+\ln s+b_0^+ & s>s_0\\
a_0^-\ln s+b_0^- & s<s_0
\end{array}\right.~,
\eq 
and it has to satisfy the boundary conditions that $v_0$ should be
continuous at $s=s_0$, $a_0^+\ln s_0+b_0^+=a_0^-\ln s_0+b_0^-$, and its
derivative discontinuous due to the Dirac's delta in (\ref{lf}),
$a_0^+/s_0-a_0^-/s_0=-1/s_0$. Summing explicitly the Fourier series
(\ref{fs}) and requiring additionally that the Coulomb
potential $v_{Coul}(s_1,s_2)$ be symmetric under exchange of $1$ and
$2$ we find 
\bq
v_{Coul}(s,\varphi;s_0,\varphi_0)=
-\ln\frac{|z-z_0|}{h(s,s_0)}+a~,
\eq
with $h(s,s_0)=1$ or $h(s,s_0)=\sqrt{ss_0}$ and $a$ a constant. Here
if we imagine the plasma confined into a disk $\Omega_R$ of radius $R$
we can choose  
\bq
v_{Coul}(s,\varphi;s_0,\varphi_0)=
-\ln\frac{|z-z_0|}{h(s,s_0)}+b~,
\eq
with $h(s,s_0)=R$ and $b=a-\ln R$ or $h(s,s_0)=\sqrt{ss_0}$ and $b=a$
so that if we rescale all the $s$ into $\lambda s$ and $R$ into
$\lambda R$ the Coulomb potential does not change apart from an
additive constant. Imagine now we are on a
plane \cite{Jancovici81b}, then $h(s,s_0)=R$.
Then in the definition of the density (\ref{rho}) at any
temperature we can change integration variables in the numerator from
$(s_i,\varphi_i)$ to
$(x_i=s_ie^{i(\varphi_i-\varphi_1)},y_i=\varphi_i-\varphi_1)$ for
$i=2,3,\ldots,N$ with Jacobian $1$. Calling
$v_b=v_b(s/R)=\bar{\rho}\int_{\Omega_R}
v_{Coul}(s,\varphi;s^\prime,\varphi^\prime)\sqrt{g(s^\prime)}\,
ds^\prime\varphi^\prime$ the 
neutralizing background potential and $v_0$ the self energy of the
background we can write       
\bq \nonumber
\rho(s_1,\varphi_1)&=&\frac{N}{Z}e^{-\beta [v_b(s_1/R)+v_0]}\int_{\Omega_R}
\prod_{i>j\ge 2}e^{-\beta v_{Coul}(\qq_i;\qq_j)}
\prod_{k=2}^N\left(\frac{|x_k-s_1|}{R}\right)^{\beta q^2}
\times \\ 
&&e^{-\beta v_b\left(x_ke^{-iy_k}/R\right)}\sqrt{g\left(x_ke^{-iy_k}\right)}\,dx_kdy_k ~.
\eq
The integral does not depend on $\varphi_1$ so
$\rho(s_1,\varphi_1)=\rho(s_1)$. Now we can make a change of variables
where $s_k\to s_k/s_1$ for $k=2,3,\ldots, N$ and $R/s_1 \to T$ so that 
\bq \nonumber
\rho(s_1)&=&\frac{N}{Z}e^{-\beta [v_b(1/T)+v_0]}\int_{\Omega_{T}}
\prod_{i>j\ge 2}e^{-\beta v_{Coul}(\qq_i;\qq_j)}
\prod_{k=2}^N\left(\frac{|x_k-1|}{T}\right)^{\beta q^2}
\times \\ \label{rho-plane}
&&e^{-\beta v_b(s_k/T)}\sqrt{g\left(s_ks_1\right)}\,s_1^{N-1}dx_kdy_k ~.
\eq
On a plane $\sqrt{g(ss_1)}=ss_1$ so that in
Eq. (\ref{rho-plane}) there is a multiplicative factor 
$s_1^{2(N-1)}$. So in the thermodynamic limit $T\to\infty$ and
$N\to\infty$ we can say that $\rho(s_1)=$constant since we know that
we must have a well defined thermodynamic limit. The same conclusion
holds on a pseudosphere (see Sec. 4.3.2 of Ref. \cite{Fantoni03jsp}),
on a cylinder (see Eq. (12a) of Ref. \cite{Choquard83}), and on a
Flamm's paraboloid (see Sec. 4.2.4 of Ref. \cite{Fantoni2008}). In
these cases the explicit analytic expression of the density has been
determined for the finite system as a function of the properties of
the surface at the special value of the coupling constant $\beta
q^2=2$. To the best of our knowledge there aren' t any analytical
results, in the literature, where the OCP has been found to have a
non-constant number density in the thermodynamic limit on a given
curved surface, probably one has to resort to numerical simulations
\cite{Caillol1982}.    
It certainly has to be expected that in a general curved surface the
OCP in the thermodynamic limit may have a non-constant density
otherwise it would mean that an OCP in the plane has a uniform density
for an arbitrary external field. It might actually be true that the
effects of the metric and the background potential cancel one another
when the potential is determined by Poisson's equation, but if it's
true, it will be necessary to solve for the potential in more detail
to prove it. 

%%%%%%%%%%%%%%%%%%%%%%%%%%%%%%%%%%%%%%%%%%%%%%%%%%%%%%%%%%%%%%%%%%%%%%%%%%%%%%
\section{The quantum case} 
%%%%%%%%%%%%%%%%%%%%%%%%%%%%%%%%%%%%%%%%%%%%%%%%%%%%%%%%%%%%%%%%%%%%%%%%%%%%%%
\label{sec:quantum}

For the quantum fluid we find for the canonical ensemble
distinguishable density matrix (the full density matrix for a system
of Bosons or Fermions is then obtained by symmetrization or
anti-symmetrization respectively) \cite{Schulman}
\bq \nonumber
\rho_D({\QQ^\prime},\QQ;\beta)=\int&&
\rho_D({\QQ^\prime},\QQ((M-1)\tau);\tau)
\cdots
\rho_D(\QQ(\tau),\QQ;\tau)\times\\
&&\prod_{j=1}^{M-1}\sqrt{\tilde{g}_{(j)}}\prod_{\alpha=1}^{2N}
\,d{Q}^{\alpha}(j\tau)~,
\eq
where as usual we discretize the imaginary time in bits
$\tau=\hbar\beta/M$ and $\QQ=(\qq_1,\ldots,\qq_N)$ with
\bq
\tilde{g}_{(i)}&=&\det||\tilde{g}_{\mu\nu}(\QQ(i\tau))||~,\\
\tilde{g}_{\mu\nu}(\QQ)&=&g_{\alpha_1\beta_1}(\qq_1)\otimes\ldots
\otimes g_{\alpha_N\beta_N}(\qq_N)~,
\eq
to get to the path integral formulation and in the small $\tau$ limit
for particles of unitary mass follows 
\bq \nonumber
\rho(\QQ(2\tau),\QQ(\tau);\tau)=(2\pi\hbar)^{-N}
\tilde{g}_{(2)}^{-1/4}\sqrt{D(\QQ(2\tau),\QQ(\tau);\tau)}
\tilde{g}_{(1)}^{-1/4}\times\\e^{\hbar\tau R(\QQ(\tau))/12}
e^{-\frac{1}{\hbar}S(\QQ(2\tau),\QQ(\tau);\tau)}~,
\eq
where $R$ is the scalar curvature of the surface, $S$ the action and
$D$ the van Vleck's determinant
\bq
D_{\mu\nu}&=&-\frac{\partial^2S(\QQ(2\tau),\QQ(\tau);\tau)}
{\partial {Q}^\mu(2\tau)\partial {Q}^\nu(\tau)}~,\\
D(\QQ(2\tau),\QQ(\tau);\tau)&=&\det||D_{\mu\nu}||~.
\eq
For example for free particles
\bq
{\cal H}&=&\frac{1}{2}\sum_{i=1}^N
g^{\alpha_i\beta_i}(\qq_i)p_{\alpha_i}p_{\beta_i}=
\frac{1}{2}\sum_{i=1}^N
g_{\alpha_i\beta_i}(\qq_i)\dot{q}^{\alpha_i}
\dot{q}^{\beta_i}~,\\
S(\QQ(2\tau),\QQ(\tau);\tau)&=&K(\QQ(2\tau),\QQ(\tau);\tau)
=\frac{1}{2}\sum_{i=1}^Nd^2(\qq_i(2\tau),\qq_i(\tau))/\tau~,
\eq
and for the fluid
\bq
S(\QQ(2\tau),\QQ(\tau);\tau)=K(\QQ(2\tau),\QQ(\tau);\tau)+
\tau V(\QQ(\tau))~.
\eq

We then find the partition function through the integral
\bq
Z=\int \rho_D(\QQ,\QQ;\beta)\,\sqrt{\tilde{g}}d\QQ~,
\eq
and the number density by
\bq
\sqrt{g(\qq_1)}\rho(\qq_1)=N\frac{\int\rho_D(\QQ,\QQ;\beta)\,
\sqrt{\tilde{g}}\prod_{i=2}^Nd\qq_i}{Z}~.
\eq
It is then apparent that by choosing the same isometry on each
imaginary time slice we reach the same conclusion of Section
\ref{sec:curvature} as for the classical (non-quantum) fluid. 

%%%%%%%%%%%%%%%%%%%%%%%%%%%%%%%%%%%%%%%%%%%%%%%%%%%%%%%%%%%%%%%%%%%%%%%%%%%%%%
\section{Conclusions} 
%%%%%%%%%%%%%%%%%%%%%%%%%%%%%%%%%%%%%%%%%%%%%%%%%%%%%%%%%%%%%%%%%%%%%%%%%%%%%%
\label{sec:conclusions}

We showed that in a surface of constant curvature without boundaries
the local number density $\rho(\qq)$ of 
a non-ideal, $(V\neq 0)$, fluid is a constant in the thermodynamic
limit. Clearly the ideal gas has constant density on every surface
regardless of the curvature and of the thermodynamic limit.   

The Coulomb potential for particles living in the surface depends on
the metric tensor and is in general a function of the geodesic
distance between the two charges. The {\sl Coulomb fluid} density is a 
constant in the thermodynamic limit in the plane \cite{Jancovici81b}
the sphere \cite{Caillol81} (and the pseudosphere
\cite{Jancovici98,Fantoni03jsp,Jancovici04}), surfaces of constant
curvatures, but also on the Flamm's paraboloid \cite{Fantoni2008}, a
surface of non-constant curvature.

We proposed a formulation for the number density which gives to the
curvature of a surface with a conformal metric (the sphere the
pseudosphere and the Flamm's paraboloid are three surfaces of this
kind) a physical interpretation as an additional external ``force''
acting on the system of particles moving in the corresponding ``flat
space''. The formulation although suggestive partly masks the
intuition of the properties of the density because of the 
fact that the pair potential is inherently related to the properties
of the curved surface, {\sl i. e.} the geodesic distance between two
points, which cannot be translated in terms of the properties
of the corresponding fluid moving in the ``flat space'' in a
straightforward way. On the other hand the formulation suggests that
the combination $\sqrt{g}\rho$ is a more fundamental quantity than
just $\rho$ itself. One can show both for the non-quantum and the
quantum fluid that if $\partial_{q^\alpha}$ is a Killing vector field of the 
surface then if we can neglect surface effects
$[\sqrt{g(\qq)}\rho(\qq)]_{,\alpha}=0$ and if 
$[\sqrt{g(\qq)}\rho(\qq)]_{,\alpha}=0$, $\forall~v$ then
$g_{,\alpha}=0$. These are the main results of our discussion. We can
also say that $g_{,\alpha}=0$ {\sl if and only if}
$[\sqrt{g(\qq)}\rho(\qq)]_{,\alpha}=0$, $\forall~v$. 

The total potential energy of the fluid moving in the ``flat space'' is
$U(\QQ)=V(\QQ)+\sum_i\phi(\qq_i;\beta)$ where the
functional dependence on $\QQ$ of the first term depends both on the
fluid model, through $v(d(\qq_i,\qq_j))$, and the kind of surface,
through $d$, whereas the functional form of the  
second term depends {\sl only} on the kind of surface. It is then to
be expected that given a fluid model the density can be non-constant
on certain surfaces. 

The OCP has uniform density in
the cylinder (see Eq. (12a) of Ref. \cite{Choquard83}), in the
pseudosphere (see Sec. 4.3.2 of Ref. \cite{Fantoni03jsp}), and in 
the Flamm's paraboloid (see Sec. 4.2.4 of
Ref. \cite{Fantoni2008}). In these cases the explicit expression of
the density has been determined for the finite system as a function of
the properties of the 
surface at the special value of the coupling constant $\beta
q^2=2$. To the best of our knowledge there aren' t any analytical
results, in the literature, where the OCP has been found to have a
non-constant number density in the thermodynamic limit on a given
curved surface, probably one has to resort to numerical simulations
\cite{Caillol1982}.    

It would be important, in the future, to be able to understand if the
surface effects on the finite system have some influence in the
conclusion that if $\partial_{q^\alpha}$ is a Killing vector field of the 
surface then $[\sqrt{g(\qq)}\rho(\qq)]_{,\alpha}=0$ in the
thermodynamic limit.

%\appendix
%%%%%%%%%%%%%%%%%%%%%%%%%%%%%%%%%%%%%%%%%%%%%%%%%%%%%%%%%%%%%%%%%%%%%%%%%%%%%%
%\section{...} 
%%%%%%%%%%%%%%%%%%%%%%%%%%%%%%%%%%%%%%%%%%%%%%%%%%%%%%%%%%%%%%%%%%%%%%%%%%%%%%
%\label{app:1}

%%%%%%%%%%%%%%%%%%%%%%%%%%%%%%%%%%%%%%%%%%%%%%%%%%%%%%%%%%%%%%%%%%%%%%%%%%%%%% 
%\begin{acknowledgements}
\ack
I would like to thank the National Institute for Theoretical Physics
of South Africa and the Institute of Theoretical Physics of the
University of Stellenbosch where the work has been started. And Karl
M\"oller for stimulating the birth of the same. 
%\end{acknowledgements}
%%%%%%%%%%%%%%%%%%%%%%%%%%%%%%%%%%%%%%%%%%%%%%%%%%%%%%%%%%%%%%%%%%%%%%%%%%%%%%
%\bibliographystyle{spbasic}
%\bibliographystyle{apsrev}
\section*{References}
\bibliographystyle{unsrt}
\bibliography{tcp-bis}

%%%%%%%%%%%%%%%%%%%%%%%%%%%%%%%%%%%%%%%%%%%%%%%%%%%%%%%%%%%%%%%%%%%%%%%%%%%%%%
%%%%%%%%%%%%%%%%%%%%%%%%%%%%%%%%%%%%%%%%%%%%%%%%%%%%%%%%%%%%%%%%%%%%%%%%%%%%%%
%%%%%%%%%%%%%%%%%%%%%%%%%%%%%%%%%%%%%%%%%%%%%%%%%%%%%%%%%%%%%%%%%%%%%%%%%%%%%%
\end{document}